\documentclass[letterpaper,twoside]{article}
\usepackage{amssymb}
\usepackage{amsmath}
\usepackage{latexsym}
\usepackage{verbatim}
\usepackage{epsfig}
\usepackage{fancyhdr}
\usepackage{pstricks}
\usepackage{pstricks-add}
\usepackage[english]{babel}

\newcommand{\papertitle}{%
From Bargmann's superselection rule to quantum\\ Newtonian spacetime}
\newcommand{\runningtitle}{%
From Bargmann's superselection rule to quantum Newtonian spacetime}
\newcommand{\pauthor}{%
H.{} Hernandez-Coronado
}
\newcommand{\paperauthor}{%
H.{} Hernandez-Coronado%
}
\begin{document}
\begin{titlepage}
\vspace*{-1cm}
\begin{flushright}
\textsf{}
\\
\mbox{}
\\

\end{flushright}
\renewcommand{\thefootnote}{\fnsymbol{footnote}}

\begin{flushright}
\textsf{\today}
\end{flushright}
\vspace{3cm}

\begin{LARGE}
\noindent\bfseries{\sffamily \papertitle}
\end{LARGE}

\noindent \rule{\textwidth}{.5mm}

\vspace*{1.6cm}

\noindent 
\begin{large}%
\textsf{\bfseries\pauthor}
\end{large}

\vspace*{.1cm}

\phantom{XX}
\begin{minipage}{1\textwidth}
\begin{it}
\noindent 
Instituto Mexicano del Petr\'oleo\\
Eje central L\'azaro C\'ardenas 152, 07730, M\'exico D.F., M\'exico.\\
\end{it}
\texttt{hhernand@imp.mx
\phantom{X}}
\end{minipage}

\vspace*{3cm}
\noindent
\textsc{\large Abstract:}
\\

Bargmann's superselection rule, which forbids the existence of superpositions of states with different mass and, therefore, implies the impossibility of describing unstable particles in non-relativistic quantum mechanics, arises as a consequence of demanding Galilean covariance of Schr\"odinger's equation. However, the usual Galilean transformations inadequately describe the symmetries of non-relativistic quantum mechanics since they fail to take into account relativistic time contraction effects which can produce non-relativistic phases in the wavefunction. In this paper we describe the incompatibility between Bargmann's rule and Lorentz transformations in the low-velocities limit, we analyze its classical origin and we show that the Extended Galilei group characterizes better the symmetries of the theory. Furthermore, we claim that a proper description of non-relativistic quantum mechanics requires a modification of the notion of spacetime in the corresponding limit, which is noticeable only for quantum particles.

\end{titlepage}
\setcounter{footnote}{0}
\renewcommand{\thefootnote}{\arabic{footnote}}
\setcounter{page}{2}
\noindent \rule{\textwidth}{.5mm}

\tableofcontents

\noindent \rule{\textwidth}{.5mm}\\

\section{Introduction}\label{I}

It is an established fact that Schr\"odinger's equation is not invariant under Galilean boosts, but it is invariant up-to-a-phase only. As a consequence a unitary representation of the Galilei group in the Hilbert space of a non-relativistic particle does not exist, but a projective one does \cite{Az,LL}. This projective representation can be regarded as a true unitary representation of the Extended Galilei group, a central extension of the Galilei one which contains the mass as a generator. The nonexistence of a unitary representation of the Galilei group implies that the transformation composed by a translation (by $\mathbf{a}$) and then a boost (by $\mathbf{v}$) followed by a translation (by $-\mathbf{a}$) and finally a boost (by $-\mathbf{v}$) to return to the original inertial system, is represented by a phase that depends on the mass of the particle. Therefore, when acting on a superposition of states with different mass, this transformation produces a non trivial interference term, while it acts as the identity on spacetime coordinates. In a few words, as Anandan put it, the origin of the latter inconsistency comes from the fact that the symmetry group of non-relativistic quantum mechanics (Extended Galilei group) is different to the spacetime's one (Galilei group) \cite{Ana}. Until recently, it was widely believed that the phase associated with the previous transformation was physically meaningless and the way out of this inconsistency consisted in imposing Bargmann's superselection rule (SSR), {\it i.e.}, forbidding superposition of states with different mass \cite{Bar}.

However, the interference term giving rise to Bargmann's rule can be shown to describe physical effects, namely, it is related to the relativistic time difference between the corresponding inertial observers to second order in $v/c$. Moreover, as pointed out by Greenberger, Bargmann's solution is unsatisfactory because since mass and energy play an equivalent role in the non-relativistic (NR) limit, superposition of states with different energy -- and therefore with different mass -- necessarily appear in non relativistic quantum mechanics \cite{Gren}. Consider for example a particle of mass $m$ that is at rest decaying into one particle of mass $m_0$ and a photon, moving along the $x$-direction. One may think of $m$ as the mass of an atom in an excited state decaying to its ground state of mass $m_0$, while emitting a photon of frequency $\omega$. Classically, photons do not exist, but this is a situation that occurs frequently in NR Quantum Mechanics. This is the case, {\it e.g.}, of the system described recently in a very debated paper \cite{Mu10}. In this case, NR momentum conservation in the $x$-direction implies that $m_0v=\hbar k$. In a reference frame moving in the $-y$-direction with velocity $u$, momentum conservation in $y$-direction gives $mu=m_0 u+\hbar ku/c$. Combining both expressions we get $m=m_0(1+v/c)$. Therefore, even in the non-relativistic case the equivalence between mass and energy shows up. And then, according to Greenberger, ``The correct way around this problem [that in NRQM, phases leading to Bargmann's SSR cannot be interpreted within the theory] is to concede that quantum mechanically, we must keep track of rest-mass and proper time differences when they appear as non vanishing phases in the NR limit and learn to incorporate them into the theory'' \cite{Gren}. Such a solution suggests that both, mass and proper time, should be treated as operators, the former being the generator of the latter \cite{Gren10a}.

Furthermore, Giulini has remarked that it does not make sense to impose a superselection rule on a parameter, and so for the mass to satisfy a SSR, it has to be dynamical and the conjugate momentum of a canonical variable \cite{Giu95}. 

This paper is an attempt to provide a minimal extension of NRQM such that superposition of mass eigenstates can be consistently described and the issues described in the previous paragraphs do not take place. Of course, a proper description of a system where superposition of states with different mass occur can be done in the framework of Relativistic Quantum Mechanics or within Quantum Field Theory, where Bargmann's SSR does not happen, and then we can take the corresponding limit when interested in non-relativistic states. Nevertheless, a NR quantum mechanical formulation of it is desirable for the sake of the theory's coherence and it is not always clear how to take such limits. The structure of the rest of the paper is as follows. In section \ref{GG}, we recall the structure of the Galilei group and how the Newtonian spacetime coordinates transform under it. We also discuss how its action on the Hilbert space of a non-relativistic quantum particle gives rise to a projective representation and we claim that the phase associated with such a representation is, in fact, a relativistic remnant that is observable. In section \ref{BSR}, we briefly describe Bargmann's superselection rule while in section \ref{CGT} we interpret Galilean transformations as canonical ones at the classical level, we relate them to the projective representation found in section \ref{GG} and we show that the appropriate group describing the symmetries of spacetime is not the Galilei group but the extended Galilei one \cite{Az}. In section \ref{EGT}, we introduce the extended Galilei group which admits a unitary representation on the corresponding Hilbert space and we show that it reproduces consistently the low velocities limit of the Poincar\'e group. Finally, in section \ref{QNS} we propose a non standard action of the Extended Galilei group on spacetime coordinates in a consistent fashion with its action on the Hilbert space, so that superposition of states with different mass are well described in non-relativistic quantum mechanics. A short conclusion is given in section \ref{DC}.

\section{Galilei group}\label{GG}

The proper Galilei group $\mathcal{G}$ consists of translations in time and space, pure Galilean transformations and rotations, generated by $H$, $\mathbf{P}$, $\mathbf{C}$ and $\mathbf{J}$, respectively. The corresponding algebra $\mathfrak{g}$ is given by:
\begin{eqnarray}
&&[P_i,P_j]=0,\hspace{1cm}[J_i,P_j]=\epsilon_{ij}^{\;\;k}P_k,\hspace{1cm}[J_i,J_j]=\epsilon_{ij}^{\;\;k}J_k,\nonumber\\
&&[C_i,C_j]=0,\hspace{1cm}[J_i,C_j]=\epsilon_{ij}^{\;\;k}C_k,\hspace{1cm}[C_i,P_j]=0,\label{Galg}\\
&&[P_i,H]=0,\hspace{1.1cm}[J_i,H]=0,\hspace{1.8cm}[C_i,H]=P_i,\nonumber
\end{eqnarray}
where Latin indexes label spacial coordinates, {\it i.e.}, $i,j,k=1,2,3$. In classical mechanics $[\cdot,\cdot]$ represent Poisson brackets while in the quantum case commutators times $(i\hbar)^{-1}$.

A general group element can be represented by $g=(b,\mathbf{a},\mathbf{v}, R)\in\mathcal{G}$, where $b$ is a real number, $\mathbf{a}$ and $\mathbf{v}$ are three dimensional real vectors and $R$ is a real $3\times 3$ orthogonal matrix. The group composition law corresponds to $g\cdot g'=(b+b',\mathbf{a}+R\mathbf{a}'+b'\mathbf{v},\mathbf{v}+R\mathbf{v}',RR')$. The identity element is $e=(0,\mathbf{0},\mathbf{0},I)$, where $I$ is the $3\times 3$ identity matrix, and the inverse element of $g=(b,\mathbf{a},\mathbf{v}, R)$ is $g^{-1}=(-b,-R^{-1}(\mathbf{a}-b\mathbf{v}),-R^{-1}\mathbf{v},R^{-1})$. Without loss of generality we will restrict ourselves to the subset $\mathcal{G}_I=\{g\in\mathcal{G}|g=(b,\mathbf{a},\mathbf{v},I)\}$, {\it i.e.}, the group elements generated by all but rotation generators $\mathbf{J}$, which clearly is a subgroup of $\mathcal{G}$. In what follows, we will make no reference to $I$ in such elements, {\it i.e.}, $\forall g\in\mathcal{G}_I$, $g=(b,\mathbf{a},\mathbf{v})$.

The action of the group $\mathcal{G}$ on Newtonian spacetime is given by:
\begin{equation}
 g\triangleright(\mathbf{x},t)\mapsto(\mathbf{x}',t')=(R\mathbf{x}+\mathbf{v}t+\mathbf{a},t+b),\label{actGTx}
\end{equation}
while the action of the subgroup $\mathcal{G}_I$ on it yields:
\begin{equation}
 g\triangleright(\mathbf{x},t)\mapsto(\mathbf{x}',t')=(\mathbf{x}+\mathbf{v}t+\mathbf{a},t+b).\label{actGTIx}
\end{equation}

\subsection{Action of $\mathcal{G}$ on Hilbert space}\label{GCSE}

It is well known that Schr\"odinger's equation is covariant under Galilean transformations. We will include the rest energy term in Schr\"odinger's equation because it plays an important role when superposition of states with different mass are considered, as it will become clear in the next sections. The standard argument goes as follows: let us assume that Schr\"odinger's equation for a particle in the presence of a scalar potential $V$ is valid in an inertial reference frame $S$ with coordinates ($\mathbf{x},t$), {\it i.e.},
\begin{equation}
 i\hbar\partial_t\psi(\mathbf{x},t)=\left(mc^2-\frac{\hbar^2}{2
m}\nabla^2+V(\mathbf{x},t)\right)\psi(\mathbf{x},t),\label{SchEq}
\end{equation}
and then, let us define another inertial reference frame $S'$ with coordinates $(\mathbf{x}',t')$ related to the
unprimed ones by the following Galilean transformation, 
\begin{equation}
\mathbf{x'}=\mathbf{x}-\mathbf{v} t,\hspace{1cm} t'=t,\label{xtp}
\end{equation}
which implies that $\partial_{t}'=\partial_{t}+\mathbf{v}\cdot\nabla$ and $\nabla'=\nabla$, where $\mathbf{v}$ is a constant vector (in Cartesian coordinates). Note that these expressions correspond to the quantum version of $E'=E-\mathbf{v}\cdot\mathbf{p}$ and $\mathbf{p}'=\mathbf{p}$, which are the usual expressions relating the particle's energy and momentum in two Galilean inertial frames with relative velocity $\mathbf{v}$. In the frame $S'$, Eq. (\ref{SchEq}) can be written as:
\begin{equation}
 i\hbar\partial_{t}'\psi'(\mathbf{x'},t')=\left(mc^2-\frac{\hbar^2}{2
m}\nabla'^2+V'(\mathbf{x'},t')\right)\psi'(\mathbf{x'},t'),\label{SchEqp}
\end{equation}
provided that
\begin{equation}
\psi'_{\rm{v}}(\mathbf{x'},t')=e^{i\Delta_{m}(\mathbf{x},t)/\hbar}
\psi(\mathbf{x},t),\label{psip}
\end{equation}
with
\begin{equation}
\Delta_{m}(\mathbf{x},t)=m(\mathbf{v}^2t/2-\mathbf{v}\cdot\mathbf{x}),\label{Delta_m} 
\end{equation}
and $V'(\mathbf{x'},t')=V(\mathbf{x},t)$. Relation (\ref{psip}) is the standard transformation expression for the wavefunction under a Galilean boost and it implies that the probability density is a Galilean scalar when $m\in\mathbb{R}$, although the wavefunction is not. 

As it is also well known, under a space and time translation the wavefunction transforms as:
\begin{equation}
 \psi_{\mathbf{a}}(\mathbf{x}+\mathbf{a},t)=\psi(\mathbf{x},t), \hspace{.5cm} \rm{and} \hspace{.5cm}\psi_b(\mathbf{x},t+b)=\psi(\mathbf{x},t),\label{PHqact}
\end{equation}
respectively, where $\psi_{\mathbf{a}}=e^{-i\mathbf{P}\cdot\mathbf{a}/\hbar}\psi$ and $\psi_b=e^{iH b/\hbar}\psi$. Relations (\ref{PHqact}) together with (\ref{psip}) describe how the group $\mathcal{G}_I$ acts on the Hilbert space of a non-relativistic quantum particle in a (Galilean) scalar potential. For a general element $g\in\mathcal{G}_I$, such an action can be written as:
\begin{equation}
 \psi_g(g\triangleright(\mathbf{x},t))=e^{i\Delta_{m}(\mathbf{x},t;g)/\hbar}\psi(\mathbf{x},t),\label{Ug}
\end{equation}
where we have written explicitly the dependence of $\Delta_m$ on the group element $g$ in the expression above and $\psi_g\equiv[\hat{U}(g)\psi]$. In fact, the set of elements $\hat{U}(g)$ is a projective representation of $\mathcal{G}_I$, {\it i.e.}, given $g,g',g''\in\mathcal{G}_I$, then $\hat{U}(g)\hat{U}(g')=\omega(g,g')\hat{U}(g\cdot g')$ such that $\omega$ is a complex number and it satisfies the following properties:
\begin{itemize}
 \item[a)] $|\omega(g,g')|=1$,

 \item[b)] $\omega(g,g')\omega(g\cdot g',g'')=\omega(g',g'')\omega(g,g'\cdot g'')$,

 \item[c)] $\omega(g,g')=\exp{(i(\Delta_m(gx,g')+\Delta_m(x,g)+\Delta_m(x,g\cdot g')))}$,
\end{itemize}
where property b) ensures the group associativity and $gx$ in c) stands for $g\triangleright (\mathbf{x},t)$. Projective representations are physically relevant because in quantum mechanics pure states are represented by rays in Hilbert space.

\subsection{Relativistic remnants}\label{RR}

The physical meaning of the phase (\ref{Delta_m}) cannot be understood within the non-relativistic quantum mechanics framework. As it has previously been noted \cite{Mash,HB,Gren83}, it is required to look into the relativistic case. The Klein-Gordon equation, 
\begin{equation}
\left(\square+\frac{m^2 c^2}{\hbar^2}\right)\phi(x)=0,\label{KG}
\end{equation}
with the ansatz:
\begin{equation}
\phi(x)=e^{-imc^2 t/\hbar}\varphi(\mathbf{x},t),\label{Rwf}
\end{equation}
reduces to:
\begin{equation}
-\frac{\hbar^2}{2m}\nabla^2\varphi(\mathbf{x},t)=i\hbar\partial_t \varphi(\mathbf{x},t)-\frac{\hbar^2}{2mc^2}\partial_t^2\varphi(\mathbf{x},t),\label{KGlimEq}
\end{equation}
and then by neglecting the last term in the r.h.s. of the above equation in the non-relativistic limit -- since it is of $\mathcal{O}\left(1/c^4\right)$--, the Schr\"odinger equation is obtained. Now, since in any other Lorentz inertial frame the wavefunction $\phi'(x')=\phi(x)$ satisfies the Klein-Gordon equation (\ref{KG}) (in primed coordinates), it follows from (\ref{Rwf}) that the $\varphi$ wavefunctions for two Lorentz inertial observers are related by:
\[
\varphi'(\mathbf{x}',t')=e^{imc^2(t'-t)/\hbar}\varphi(\mathbf{x},t),
\]
where $t'-t=(\mathbf{v}^2t/2-\mathbf{v}\cdot\mathbf{x})/c^2+\mathcal{O}(1/c^4)$, such that in the non-relativistic limit, the previous expression reduces precisely to
\begin{equation}
\varphi'(\mathbf{x}',t')=e^{im(\mathbf{v}^2t/2-\mathbf{v}\cdot\mathbf{x})/\hbar}\varphi(\mathbf{x},t),\label{boostLim}
\end{equation}
which agrees with relations (\ref{psip}) and (\ref{Delta_m}). Thus, the phase can be identified as some relativistic remnant: it is due to the time difference of order $1/c^2$ between the different inertial observers.

\begin{figure}[!tb]
\centering
\includegraphics[scale=1]{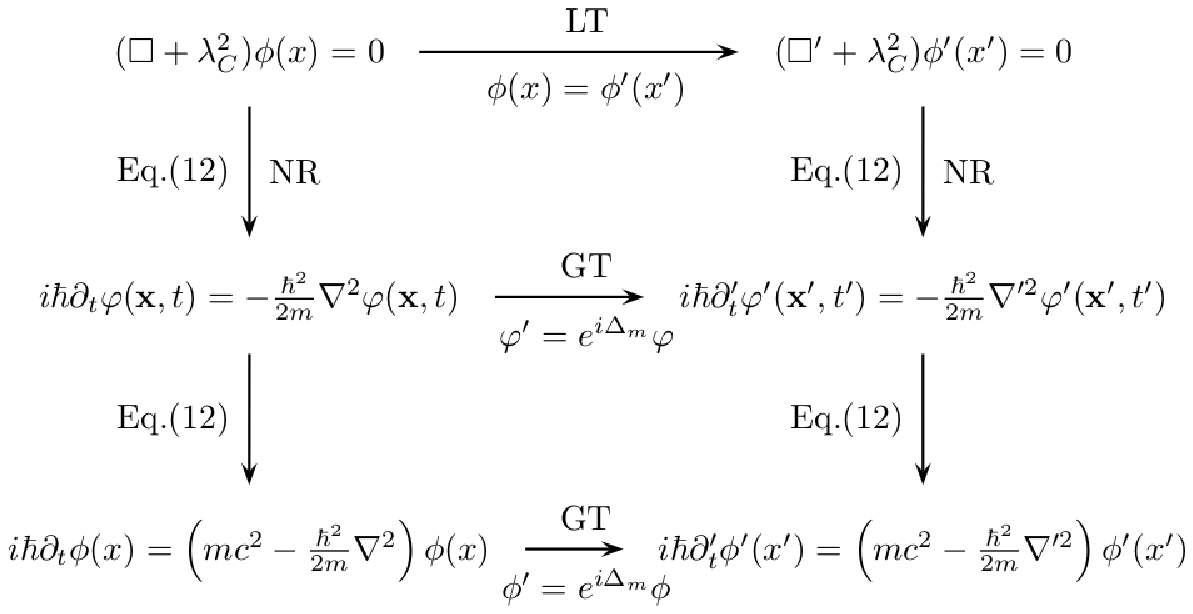}
\caption{Diagrammatic description of the incompatibility between Galilean inertial observers and non-relativistic Lorentz ones. While the invariance of the Klein-Gordon equation under Lorentz transformations (LT) allows for $\phi(x)$ to be a Lorentz scalar even in the NR limit, the covariance of the Schr\"odinger equation under Galilean transformations (GT) implies that $\phi(x)$ is not. $\lambda_C=mc/\hbar$ is the particle's Compton wavelength.}
\label{Fig.1}
\end{figure}

At this point some remarks are in order. The interpretation of the phase (\ref{Delta_m}) given in the previous analysis is based on Lorentzian -- rather than on Galilean -- symmetry arguments, and since it has a relativistic origin, the phase $\Delta_m$ cannot be understood within the non-relativistic quantum mechanics theory. On the other hand, the ansatz (\ref{Rwf}), allows us to identify the non-relativistic terms (because the order in $c$ of a differential operator depends on the functions it is applied to). That is, for a state of the form $\phi=e^{-imc^2t/\hbar}\varphi$ that fulfills the condition $|(i\hbar \partial_t-mc^2)\phi|=|i\hbar\partial_t\varphi|\ll mc^2|\phi|$ in a given Lorentz reference frame, there exists a class of Lorentz inertial observers moving with relative velocity $v\ll c$ (NR inertial observers) for which such a condition holds. Then, the term $\hbar^2\partial_t^2\varphi/(2mc^2)$ is negligible in the NR limit for all such observers, and they all will describe the particle's state by the corresponding non-relativistic wavefunction $\varphi$ satisfying the standard Schr\"odinger equation (without the rest energy term). Now, we can write $\varphi=e^{imc^2t/\hbar}\phi$ back in the equation so obtained, such that $\phi$ obeys the Schr\"odinger equation with the rest energy term (\ref{SchEq}) for all NR inertial observers. Since $\phi$ is Lorentz invariant, we would expect that for any two NR inertial observers their wavefunctions would be related as $\phi'(x')=\phi(x)$. However, as we have shown, Galilean covariance of equation (\ref{SchEq}) implies that any two Galilean observers would describe the particle's state by two wavefunctions related as $\phi'(x')=e^{i\Delta_m/\hbar}\phi(x)$, which is in contradiction with their Lorentz invariance (see Fig.\ref{Fig.1}). This indicates that Galilean inertial observers do not agree with NR inertial ones, and this suggests that the Galilei group does not describe the symmetry of the non-relativistic spacetime as we will confirm in section \ref{CGT}. A hint that supports the previous statement is that the spacetime metric that is invariant under Galilei transformations is $c^2\eta_{\mu\nu}=\rm{diag}(0,0,0,-c^2)$, which describes the Newtonian spacetime (parametrized by coordinates $x^{\mu}=(\mathbf{x},t)$) only in the strict limit $c\rightarrow\infty$. Greek indexes label spacetime coordinates, {\it i.e.}, $\mu,\nu=1,2,3,4$, with $x^4=t$.

\section{Bargmann's superselection rule}\label{BSR}

The discussion of the previous section holds formally for $m\rightarrow\tilde{m}\in\mathbb{C}$, which presumably would describe an unstable particle (in certain regime) \cite{MS}. However, in this case the wavefunctions of the two inertial observers are related by:
\begin{equation}
|\tilde{\psi}|^2=e^{-2\rm{Im}{(\Delta_{\tilde{m}})}/\hbar}|\psi|^2\neq|\psi|^2,\label{decTwf}
\end{equation}
and therefore it is not clear that the same physical interpretation can be given to the ``Galilean transformation'' thus defined. This result is related to Bargmann's superselection rule.

In order to see this, we will reproduce here Bargmann's original argument \cite{Bar}. Suppose $\psi(\mathbf{x},t)=\psi_1(m_1,\mathbf{x},t)+\psi_2(m_2,\mathbf{x},t)$ is a superposition of states with two masses in an inertial reference frame $S$. Now let us perform the following composed transformation:
\begin{itemize}
 \item[i)] $g_{1}=(0,\mathbf{a},\mathbf{0})$: a translation by $\mathbf{a}$. 
\[
\psi_{g_1}(\mathbf{x}+\mathbf{a},t)=\psi(\mathbf{x},t)=\psi_1(\mathbf{x},t)+\psi_2(\mathbf{x},t),
\]

  \item[ii)] $g_{2}=(0,\mathbf{0},\mathbf{v})$: a pure boost by $\mathbf{v}$.
 \begin{eqnarray*}
\psi_{g_2\cdot g_1}(\mathbf{x}+\mathbf{a}+\mathbf{v}t,t)&=&e^{i\Delta_{m_1}(\mathbf{x}+\mathbf{a},t;g_2)/\hbar}\psi_{1}(\mathbf{x},t)+e^{i\Delta_{m_2}(\mathbf{x}+\mathbf{a},t;g_2)/\hbar}\psi_{2}(\mathbf{x},t),
  \end{eqnarray*}
 \item[iii)] $g_{3}=(0,-\mathbf{a},\mathbf{0})=g_1^{-1}$: a translation by $-\mathbf{a}$. 
\begin{eqnarray*}
\psi_{g_3\cdot g_2\cdot g_1}(\mathbf{x}+\mathbf{v}t,t)&=&e^{i\Delta_{m_1}(\mathbf{x}+\mathbf{a},t;g_2)/\hbar}\psi_{1}(\mathbf{x},t)+e^{i\Delta_{m_2}(\mathbf{x}+\mathbf{a},t;g_2)/\hbar}\psi_{2}(\mathbf{x},t),
\end{eqnarray*}
 
  \item[iv)] $g_{4}=(0,\mathbf{0},\mathbf{-v})=g_2^{-1}$: a pure boost by $-\mathbf{v}$. 
 \begin{eqnarray}
\psi_{g_4\cdot g_3\cdot g_2\cdot g_1}(\mathbf{x},t)&=&e^{im_1 \mathbf{v}\cdot\mathbf{a}/\hbar}\psi_1(\mathbf{x},t)+e^{im_2 \mathbf{v}\cdot\mathbf{a}/\hbar}\psi_2(\mathbf{x},t),\label{psipp}
\end{eqnarray}

\end{itemize}
where $m \mathbf{v}\cdot\mathbf{a}=\Delta_{m}(\mathbf{x}+\mathbf{a},t;g_2)+\Delta_{m}(\mathbf{x}+\mathbf{v}t,t;g_4)$. Now, since $g_{4}\cdot g_{3}\cdot g_{2}\cdot g_{1}=e$, and then $\hat{U}(g_{4})\hat{U}(g_{3})\hat{U}(g_{2})\hat{U}(g_{1})=\tilde{\omega}(g_{1},g_{2},g_{3},g_{4})\hat{U}(e)$  we expect that
\begin{equation}
\psi_{g_4\cdot g_3\cdot g_2\cdot g_1}(\mathbf{x},t)=\tilde{\omega}(g_{1},g_{2},g_{3},g_{4})\psi(\mathbf{x},t),
\end{equation}
where $\tilde{\omega}=\omega(g_{4},g_{3})\omega(g_{2},g_{1})\omega(g_{4}\cdot g_{3},g_{2}\cdot g_{1})$ such that $|\tilde{\omega}|=1$. But as it can be observed from expression (\ref{psipp}), this occurs only if Bargmann's superselection rule is imposed, {\it i.e.}, $m_2=m_1$. 

As it has been shown both effects, the exponential term in relation (\ref{decTwf}) and the non trivial phase in relation (\ref{psipp}), have the same origin, they are due to the time dilation effect between the corresponding observers.

\section{Galilean as canonical Transformations}\label{CGT}

Since in the relativistic case, the corresponding (Klein-Gordon or Dirac) equation is invariant under the Poincar\'e group, and then the wavefunction is truly a scalar, there is not a relativistic analogue of Bargmann's superselection rule (which is ``obviously approximate since Nature is relativistic.'' \cite{Az}). Clearly, this rule has its origin in that the wavefunction is not invariant under $\mathcal{G}_I$ but only up-to-a-phase invariant and it can be traced back to the classical regime. 

The non-relativistic dispersion relation $E=mc^2+\mathbf{p}^2/2m$ is not invariant under the Galilean transformation (\ref{xtp}), that is,
\begin{equation}
 E=mc^2+\frac{\mathbf{p}^2}{2m} \hspace{.5cm}\rightarrow\hspace{.5cm}E'=mc^2+\frac{\mathbf{p}'^2}{2m}-\mathbf{v}\cdot\mathbf{p}',\label{NRdispersion}
\end{equation}
however, it takes the standard form when a further canonical transformation, generated by the type $3$ function 
\begin{equation}
F_{3}(\mathbf{p},\mathbf{Q},t)=-\mathbf{Q}\cdot(\mathbf{p}-m\mathbf{v})+\frac{1}{2}m\mathbf{v}^2 t,\label{Fcan}
\end{equation}
is performed. Such a transformation, nevertheless, depends on the mass of the particle. 
The above function generates the following canonical transformation:
\begin{eqnarray}
&&\mathbf{q}=-\frac{\partial F_3}{\partial \mathbf{p}}=\mathbf{Q},\label{Fq}\\
&&\mathbf{P}=-\frac{\partial F_3}{\partial \mathbf{Q}}=\mathbf{p}-m\mathbf{v},\label{FP}\\
&&K=H+\frac{\partial F_3}{\partial t}=H+\frac{1}{2}m\mathbf{v}^2.\label{FK}
\end{eqnarray}
Not surprisingly, also the Galilean transformation can be regarded as a canonical transformation, generated by the type $2$ function:
\begin{equation}
 F_2(\mathbf{q},\mathbf{P},t)=\mathbf{P}\cdot (\mathbf{q}-\mathbf{v}t).\label{Gcan}
\end{equation}
And therefore, the composed transformation generated by $F_3\circ F_2$ leaves the non-relativistic dispersion relation invariant, that is, 
\begin{equation}
 F_{3}\circ F_{2}:(E,\mathbf{p})\mapsto\left(E''=E-\mathbf{v}\cdot\mathbf{p}+\frac{1}{2}m\mathbf{v}^2, \mathbf{p}''=\mathbf{p}-m\mathbf{v}\right),\label{epp}
\end{equation}
such that $E''=\mathbf{p}''^2/2m+mc^2$ is satisfied provided that $E=\mathbf{p}^2/2m+mc^2$ holds.

The quantum version of canonical transformations are unitary transformations, in particular the quantum version of (\ref{Fcan}) is given by:
\begin{equation}
 U_{3}=e^{-iF_3/\hbar}=\exp{\left(-\frac{i m}{\hbar}\left(\mathbf{v}\cdot \hat{\mathbf{x}}+\frac{1}{2}\mathbf{v}^2 t\right)\right)},\label{UF}
\end{equation}
where the first term from expression (\ref{Fcan}) generates the identity transformation. And correspondingly, the quantum version of (\ref{Gcan}) takes the form:
\begin{equation}
 U_{2}=e^{-iF_2/\hbar}=\exp{\left(\frac{i \mathbf{v}t}{\hbar}\cdot \hat{\mathbf{P}}\right)}.\label{UG}
\end{equation}

The unitary transformation $U_3 U_2$, performs the quantum version of Galilean transformations considered in section \ref{GCSE} and produces the phase appearing in expression (\ref{psip}), as pointed out in previous works \cite{HB}. The latter can be seen by writing down $\psi'(x')=\langle x'|\psi'\rangle$ explicitly, with $|x'\rangle=U_2|x\rangle$ and $|\psi'\rangle= U_3 U_2|\psi\rangle$. Notice that the Galilean transformation on the spacial coordinate basis $|x\rangle$ and on the wavevectors $|\psi\rangle$  is performed by different unitary operators. This happens because the generator of $U_3$ does not change the spacial coordinates, {\it cf.} relation (\ref{Fq}). Clearly, if the complete transformation $U_3 U_2$ is applied on $|x\rangle$ as well, then the non-relativistic wavefunction is a scalar, {\it i.e.}, $\langle x|\psi\rangle=\langle x''|\psi'\rangle$, with $|x''\rangle=U_3 U_2|x\rangle$.

From the previous analysis, nonetheless, we can see that the boost\footnote{It transforms properly not only the coordinates but the momentum and energy as well, {\it cf}. (\ref{epp}).} $U_3 U_2=e^{i(\mathbf{v}t\cdot\mathbf{P}-\mathbf{v}\cdot m\mathbf{x})/\hbar}$, that corresponds to an exact symmetry of the system, is generated by $\mathbf{C}=m\mathbf{x}-\mathbf{P}t$ and satisfies the relation $[C_i,P_j]=[mx_i,P_j]=m\delta_{ij}$. It follows from the latter non vanishing commutator (compare with the corresponding relation in (\ref{Galg})) that the Galilei group cannot be the symmetry group of non-relativistic quantum mechanics.

\section{Extended Galilean Transformations}\label{EGT}

From the above discussion and as it will be shown, the symmetry group describing non-relativistic quantum mechanics -- the Extended Galilei group $\tilde{\mathcal{G}}$ -- is given by a nontrivial central extension of the Galilei group consisting of introducing another generator $M$, which belongs to the center of $\mathcal{G}$ (it commutes with any other generator) and such that the algebra $\mathfrak{g}$ remains the same except for the relation
\begin{equation}
 [C_i,P_j]= M\delta_{ij}.\label{CPM}
\end{equation}
Notice that the above commutation relation coincides with the expression we found at the end of the previous section. Since we have introduced another generator, a group element $\tilde{g}\in\tilde{\mathcal{G}}$ is labeled now by $\tilde{g}=(b_m,b,\mathbf{a},\mathbf{v},R)$ with $b_m\in\mathbb{R}$. Similarly, the subgroup $\mathcal{G}_I$ can be centrally extended to $\tilde{\mathcal{G}}_I$ such that $\tilde{g}=(b_m,b,\mathbf{a},\mathbf{v})\in\tilde{\mathcal{G}}_I$, and the group product is defined as: 
\begin{equation}
\tilde{g}\cdot\tilde{g}'=\left(b_m+b_m'+\mathbf{v}\cdot\mathbf{a}'+\frac{1}{2}\mathbf{v}^2 b',b+b',\mathbf{a}+\mathbf{a}'+b'\mathbf{v},\mathbf{v}+\mathbf{v}'\right),\label{gpExt}
\end{equation}
 for $\tilde{g},\tilde{g}'\in\tilde{\mathcal{G}}_I$. The identity element is $\tilde{e}=(0,0,\mathbf{0},\mathbf{0})$ and the inverse of $\tilde{g}$ is
\begin{equation}
\tilde{g}^{-1}=\left(-b_m+\mathbf{v}\cdot\mathbf{a}-\frac{1}{2}\mathbf{v}^2 b,-b,-(\mathbf{a}-b\mathbf{v}),-\mathbf{v}\right). \label{ginExt}
\end{equation}
In previous discussions \cite{Az,LL}, the action of the extended Galilei group $\tilde{\mathcal{G}}$ on spacetime is retained as in (\ref{actGTx}), such that for $\tilde{g}=(b_m,b,\mathbf{a},\mathbf{v})\in\tilde{\mathcal{G}}_I$,
\begin{equation}
 \tilde{g} \triangleright (\mathbf{x},t)\mapsto (\mathbf{x}',t')=(\mathbf{x}+\mathbf{v}t+\mathbf{a},t+b),\label{tactGTIx}
\end{equation}
which is independent of $b_m$ ({\it i.e.}, it is unfaithful), while its action on the Hilbert space gives rise to a unitary representation $\tilde{U}(\tilde{g})$, such that $[\tilde{U}(\tilde{g})\psi](\tilde{g}x)=\psi(x)$ with $\tilde{U}(\tilde{g})\tilde{U}(\tilde{g}')=\tilde{U}(\tilde{g}\cdot\tilde{g}')$. This unitary representation is equivalent to the projective representation $U(g)$. In particular, the Bargmann composed transformation $\tilde{g}_4\cdot\tilde{g}_3\cdot\tilde{g}_2\cdot\tilde{g}_1=(0,0,\mathbf{0},-\mathbf{v})\cdot(0,0,-\mathbf{a},\mathbf{0})\cdot(0,0,\mathbf{0},\mathbf{v})\cdot(0,0,\mathbf{a},\mathbf{0})=(\mathbf{a}\cdot\mathbf{v},0,\mathbf{0},\mathbf{0})=\tilde{g}_m$ is represented by 
\begin{equation}
 \tilde{U}(\tilde{g}_{4})\tilde{U}(\tilde{g}_{3})\tilde{U}(\tilde{g}_{2})\tilde{U}(\tilde{g}_{1})=\tilde{U}(\tilde{g}_m)=\exp{(iM\mathbf{a}\cdot\mathbf{v}/\hbar)},\label{EGTBT}
\end{equation}
which is not the group identity, while its action on spacetime yields $\tilde{g}_m\triangleright (\mathbf{x},t)=(\mathbf{x},t)$. It is argued that since $M$ commutes with all the observables, then it decomposes the Hilbert space into coherent subspaces characterized by its eigenvalue $m$ in such a way that no observable can mix these subspaces \cite{Az}. This is Bargmann's superselection rule. However, this rule cannot take into account appropriately the physical world's symmetry in the low-velocities limit, which we know is described by the Poincar\'e group, as will be discussed in what follows.  

Note that in the Poincar\'e algebra the commutator between the boost generators $\mathbf{K}$ and the spacial translations $\mathbf{P}$ is given by $[K_i,P_j]=H\delta_{ij}/c^2$, which reduces precisely to the expression (\ref{CPM}) to leading order in the non-relativistic limit. With the help of the Baker-Campbell-Hausdorff formula, let us now write the relativistic version of the transformation considered by Bargmann, up to $\mathcal{O}(1/c^2)$:
\begin{equation}
 e^{-i \mathbf{v}\cdot\mathbf{K}/\hbar}e^{-i\mathbf{a}\cdot\mathbf{P}/\hbar}e^{i \mathbf{v}\cdot\mathbf{K}/\hbar}e^{i\mathbf{a}\cdot\mathbf{P}/\hbar}=e^{iH \mathbf{a}\cdot\mathbf{v}/\hbar c^2}e^{i (\mathbf{v}\cdot{\mathbf{a}})(\mathbf{v}\cdot\mathbf{P})/2\hbar c^2},\label{PBT}
\end{equation}
and accordingly, under the above transformation, the coordinates of the event $(\mathbf{x},t)$ are given by
\begin{equation}
(\mathbf{x}',t')=\left(\mathbf{x}+\frac{(\mathbf{v}\cdot\mathbf{a}) \mathbf{v}}{2c^2},t+\frac{\mathbf{v}\cdot\mathbf{a}}{c^2}\right),\label{PBTxt} 
\end{equation}
up to $\mathcal{O}(1/c^2)$. Then, in the non-relativistic limit, expression (\ref{PBT}) reduces precisely to expression (\ref{EGTBT}). Therefore, we note that the transformation generated by $\tilde{g}_{4}\cdot \tilde{g}_{3}\cdot \tilde{g}_{2}\cdot \tilde{g}_{1}$ produces a relativistic time translation, that shows up as a non-relativistic phase when acting on the wavefunction of a non-relativistic quantum particle. As a consequence, the action of $\tilde{\mathcal{G}}$ on spacetime should not be (\ref{tactGTIx}), otherwise we would be identifying two spacetime points with different time coordinates -- that can be distinguished by a quantum particle -- as a single one\footnote{Notice that in the non-relativistic limit, quantum states $\psi$ are such that $|\mathbf{P}\psi|\ll mc|\psi|$ and therefore $|H\psi|=mc^2|\psi|+\mathcal{O}(c^0)$, that is, while $H$ can generate non-relativistic time translations even for time intervals of $\mathcal{O}(1/c^2)$, $\mathbf{P}$ produces non-relativistic spacial translations only for spacial intervals of order $c^0$. And consequently a quantum particle can only keep track of corrections of order $1/c^2$ in the time coordinate.}. Bargmann's superselection rule relies on this identification. 

The bottom line is that the phase (\ref{Delta_m}) has a physical interpretation and we can identify $M$ as the generator of time dilations of order $1/c^2$. Moreover, it has been claimed that the interference pattern observed in the COW experiment \cite{Gren83} as well as in the Sagnac effect \cite{Die90,Ana}, which both have been measured \cite{cow,csw} and can be interpreted as the quantum version of the twin paradox, can be related to the phase $\Delta_m$ appearing in expression (\ref{psip}). 

The previous description can be made in the Lagrangian formalism as well, and it will prove useful. A classical free particle Lagrangian $L=\frac{1}{2}m \dot{\mathbf{x}}^2$ is not invariant under a Galilei transformation (\ref{xtp}), {\it i.e.}, $L\rightarrow L'=\frac{1}{2} m \dot{\mathbf{x}}'^2 +d\Delta_m/dt'$. The term $\Delta_m(\mathbf{x}',t';\mathbf{v})$, which is the same as the one appearing in the relation (\ref{psip}), corresponds to a boundary term which does not introduce any physical effect when $m\in\mathbb{R}$ neither in classical nor in quantum mechanics. In the former because it does not affect Lagrange equations and in the latter because since it does not depend on the path but only on the initial and final points, it translates into a global phase in the quantum amplitude 
 \begin{equation}
\langle \mathbf{x}_f,t_f|\mathbf{x}_i,t_i\rangle=\int_{x_i}^{x_f}\mathcal{D}[x(t)]e^{\frac{i}{\hbar}S(\mathbf{x},\dot{\mathbf{x}})} \rightarrow
e^{\frac{i}{\hbar}\Delta_m(\mathbf{x}',t';\mathbf{v})}\int_{x'_i}^{x'_f}\mathcal{D}[x'(t')]e^{\frac{i}{\hbar}S'(\mathbf{x}',\dot{\mathbf{x}}')}.
\end{equation}
For a superposition of states with different mass (or unstable particles, with $m\rightarrow \tilde{m}$), the boundary term is not a simple global phase and therefore the probability density takes different values for different Galilean inertial observers. 

The above boundary term has been identified with the previously introduced central extension of the Galilei group, and it has been suggested \cite{MKS} that if Newtonian spacetime is embedded in a five dimensional space according to $X^A=(\mathbf{x},t,s)$, with $A=1, ..., 5$, such that $s$ transforms under Galilean transformations as:
\begin{equation}
s'= s-\mathbf{x}\cdot\mathbf{v}+\frac{1}{2}\mathbf{v}^2 t,\label{GTs}
\end{equation}
then the inner product defined as $(X|Y)=h_{AB}X^A Y^B=\mathbf{X}\cdot\mathbf{Y}-X_4 Y_5-X_5 Y_4$ is invariant under Galilean transformations (\ref{xtp}) and (\ref{GTs}), where $h_{AB}$ is the {\it Galilean metric}. In particular, the Lagrangian redefined as:
\begin{equation}
 L\equiv mc^2\left(\eta_{AB}+\frac{1}{2c^2} h_{AB}\right)\dot{X}^A\dot{X}^B=-mc^2\left(1-\frac{1}{c^2}\left(\frac{1}{2}\dot{\mathbf{x}}^2-\dot{s}\right)\right)\label{Lag}
\end{equation}
with $\eta_{AB}=\rm{diag}(0,0,0,-1,0)$, is a scalar in this 5-dimensional spacetime. The fifth coordinate $s$ can be interpreted as the relativistic correction to time intervals of order $1/c^2$.

Summing up, the situation is that the composed transformation considered by Bargmann is equivalent to the identity when acting on the coordinates of the Newtonian spacetime, but is not the identity when acting on the Hilbert space of a non-relativistic particle. We can assume either, that Bargmann's transformation is equivalent to the identity indeed and that the phase associated with the projective representation is unphysical, or that Bargmann's transformation is not equivalent to the identity but to a physical transformation generated by $M$. That is, in general, we can assume that the symmetry group of the physical world in the NR limit is either, the Galilei group or the Extended one. The first choice leads us to Bargmann's SSR, while the second one demands us to modify the action of the Extended Galilei group on the coordinates of Newtonian spacetime in consistency with its action on the corresponding Hilbert space. As we have shown, the action of Poincar\'e group on both, Minkowski spacetime coordinates and the relativistic wavefunctions, allows us to conclude in favor of the second choice.

\section{Quantum Newtonian spacetime}\label{QNS}

As it has been established, we need to modify the action of the Extended Galilei group on the Newtonian spacetime for consistently describing the non-relativistic quantum particle dynamics. The most direct way to obtain such an action is to expand the action of the Poincar\'e group on Minkowski spacetime to the corresponding order in $c$. However it can be shown that the above procedure cannot yield a linear action of the Extended Galilei group on spacetime coordinates up to $\mathcal{O}(1/c^2)$. The reason is that we are interested in spacetime intervals of order $1/c^2$ such that their product with the corresponding Poincar\'e generator are independent of $c$, so then we need to take into account that in the Poincar\'e algebra the boost generators along different directions produce a rotation. However such a rotation does not take place in the algebra related to the Extended Galilei group. The lack of a linear action of the Extended Galilei group on the coordinates of Newtonian spacetime is a formal problem only, and it does not have physical consequences because for spacial intervals of order $1/c^2$ all unitary transformations produce phases of $\mathcal{O}(1/c^2)$ when acting on quantum states. 

Therefore, one approach that can be followed in order to obtain the sought action, is to separate (somehow arbitrarily) the time coordinate of Minkowski spacetime in the Newtonian time $t$ (that is the same for all inertial observers) plus a correction of order $1/c^2$, $s$, and to consider both, $t$ and $s$ as independent coordinates. Such an approach is in agreement with what Giulini has claimed \cite{Giu95}, namely that for the mass to satisfy a superselection rule, it must be dynamical and, accordingly, we have to introduce its conjugate coordinate (the correction of $\mathcal{O}(1/c^2)$ to relativistic time, $s$), and also it is very well suited to the five dimensional description of Newtonian spacetime discussed in section \ref{EGT}.

\subsection{Dynamical mass}

Let us introduce $s$ as a fifth independent coordinate in spacetime, such that the action of the Extended Galilei group on $5$-dimensional spacetime coordinates takes the form:
\begin{equation}
\tilde{g}\; \triangleright_5\; (\mathbf{x},t,s) \mapsto\left(\mathbf{x}+\mathbf{v}t+\mathbf{a},t+b,s+b_m+\mathbf{v}\cdot\mathbf{x}+\frac{1}{2}\mathbf{v}^2 t\right),\label{sact}
\end{equation}
and $M$ is represented by $i\hbar\partial_s$. It is not difficult to see that $\triangleright_5$ thus defined is an action indeed (see \cite{Az,Giu95}). In this scheme the mass is the conjugate momentum of $s$ and therefore it is dynamical \cite{Giu95}, so the corresponding Hamiltonian takes the form
$H=\dot{\mathbf{x}}\cdot \mathbf{p}-m\dot{s}-L$ with $L$ given by (\ref{Lag}). At the quantum level this implies that the Schr\"odinger equation takes the form:
\begin{equation}
i\hbar\partial_t\psi(\mathbf{x},t,s)=\left(i\hbar c^2\partial_s-\frac{\hbar}{2i}\partial_s^{-1}\nabla^2+V(\mathbf{x},t,s)\right)\psi(\mathbf{x},t,s).\label{Sch5}
\end{equation}
Formally $\partial_s^{-1}$ can be defined when applied to a general function as $\partial_s^{-1} \varphi(s)=-i\hbar\int m^{-1} e^{ims/\hbar}\tilde{\varphi}_m dm$, where $\tilde{\varphi}_{m}$ is $\varphi(s)$'s Fourier transform. For a general function, the latter integral does not converge. We shall assume that the physical states describing non-relativistic particles are such that it converges. For mass eigenstates the previous equation reduces to the standard Schr\"odinger equation. And in terms of the Fourier transform $\tilde{\varphi}_m$, the previous equation coincides with the equation defined by Giulini \cite{Giu95}. 

Under a pure extended Galilean transformations, spacetime coordinates transform as:
\begin{equation}
\mathbf{x}'=\mathbf{x}+\mathbf{v} t,\hspace{1cm}
t'=t,\hspace{1cm}
s'=s+\mathbf{v}\cdot\mathbf{x}+\frac{1}{2}\mathbf{v}^2 t,\label{xtsEGT}
\end{equation}
({\it cf.} relation (\ref{GTs})) which implies the following relations:
\begin{equation}
\partial_t=\partial_{t'}+\mathbf{v}\cdot\nabla'+\frac{\mathbf{v}^2}{2}\partial_{s'},\hspace{1cm}\nabla=\nabla'+\mathbf{v}\partial_{s'},\hspace{1cm}\partial_s=\partial_{s'}.\label{qeep}
\end{equation}
Accordingly, assuming that $\psi(\mathbf{x},t,s)$ satisfies Schr\"odinger's equation (\ref{Sch5}) in the inertial reference frame $S_5$ with coordinates $(\mathbf{x},t,s)$, then $\tilde{\psi}'(\mathbf{x}',t',s')=\psi(\mathbf{x},t,s)$ satisfies the same Schr\"odinger equation in the primed inertial reference frame $S'_5$ with coordinates $(\mathbf{x}',t',s')$ given by (\ref{xtsEGT}). Therefore, in this framework there is no need for Bargmann's superselection rule, and nothing prevents us from describing consistently superposition of states with different mass. The above expressions (\ref{qeep}) correspond to the quantum version of expression (\ref{epp}), {\it i.e.}, $E=E'-\mathbf{v}\cdot\mathbf{p}'+\frac{1}{2}m\mathbf{v}^2$ and $\mathbf{p}=\mathbf{p}'-m\mathbf{v}$, which leaves the non-relativistic dispersion relation invariant just as the transformations (\ref{qeep}) leave the Schr\"odinger equation invariant. 

Furthermore, when a transformation to a uniform accelerating frame is considered, expressions (\ref{xtsEGT}) generalize to
\begin{equation}
\mathbf{x}'=\mathbf{x}+\frac{1}{2}\mathbf{g}t^2,\hspace{.5cm}t'=t,\hspace{.5cm}s'=s+\mathbf{g}t\cdot\mathbf{x}+\frac{1}{3}\mathbf{g}^2 t^3,\label{xtsAc}
\end{equation}
with $\mathbf{g}$ constant, and correspondingly relations (\ref{qeep}) to:
\begin{equation}
\partial_t=\partial_{t'}+\mathbf{g}t\cdot\nabla'+\left(\mathbf{g}\cdot\mathbf{x}+\mathbf{g}^2 t^2\right)\partial_{s'},\hspace{.5cm}\nabla=\nabla'+\mathbf{g}t\;\partial_{s'},\hspace{.5cm}\partial_s=\partial_{s'}.\label{dxtsAc}
\end{equation}
So then, if $\psi(\mathbf{x},t,s)$ satisfies the Schr\"odinger equation (\ref{Sch5}) in an inertial reference frame with coordinates $(\mathbf{x},t,s)$, with $V=0$, $\psi'(\mathbf{x}',t',s')$ satisfies:
\begin{equation}
i\hbar\partial_{t}'\psi'(\mathbf{x}',t',s')=\left(Mc^2-\frac{\hbar}{2}M^{-1}\nabla'^2-M\mathbf{g}\cdot\mathbf{x}'\right)\psi'(\mathbf{x}',t',s'),\label{SchEqG}
\end{equation}
in an accelerating frame with coordinates $(\mathbf{x}',t',s')$ given by relations (\ref{xtsAc}). The previous equation describes a quantum particle in a constant gravitational potential, in agreement with the equivalence principle. That the transformation (\ref{xtsAc}) is the appropriate one relating the coordinates of an accelerating observer to those of an inertial observer can be seen by taking the corresponding limit of the relativistic case (see \cite{Gren79}).

Finally, notice that if we separate the time as $t_{r}=t+s/c^2$, then the Minkowski line element $dl^2=-c^2dt_r^2+d\mathbf{x}^2$ can be written, up to order $1/c^2$, as $dl^2=-c^2 dt^2+d\mathbf{x}^2-2dt ds=c^2 g_{AB}dx^A dx^B$, which is invariant under extended Galilean transformations (\ref{xtsEGT}), and thus, we can recognize the {\it Galilean metric} $h_{AB}=c^2(g_{AB}-\eta_{AB})$ in expression (\ref{Lag}) as a perturbation to the metric $\eta_{AB}$, which is the $5$-dimensional version of the Galilean invariant metric $\eta_{\mu\nu}$, introduced previously in subsection \ref{RR}.

\section{Discussion and conclusions}\label{DC}

As we have discussed, a unitary representation of the Galilei group in the Hilbert space of a non-relativistic quantum particle does not exist, but a projective representation does. This projective representation leads to the formulation of Bargmann's superselection rule, but the phase associated with transformations in this projective representation contains information about physical observables. Therefore, by imposing the superselection rule we are neglecting some physically observable effects. As we have shown, the extended Galilei group is more suitable for describing the symmetry of the physical world in the low-velocities limit, which can be represented by true unitary operators in the corresponding Hilbert space. Nevertheless, in order for the symmetries described by this group to be incorporated in a consistent way into the theory, we must modify accordingly the -- traditionally unfaithful -- action of the group on the spacetime coordinates. For this purpose we need to introduce the relativistic correction to time as a fifth independent coordinate and to elevate the mass as its generator. This framework allows us to show that the Schr\"odinger equation is \emph{invariant} under the appropriate transformation law. As a consequence, Bargmann's superselection rule which forbids superposition of states with different mass is not valid, since it relies on the identification of points in spacetime with different time coordinates. Therefore, the Schr\"odinger equation with $m$ replaced by an operator $M$, can describe unstable particles effectively, indeed. In particular, the result we had previously found, according to which the probability density associated to the particle's wavefunction by two different inertial observers would differ by an exponential term ({\it cf.} equation (\ref{decTwf})), can be understood naturally: the  wavefunctions of the different observers do not describe the state of the particle at the same time, and since the particle is decaying, then the exponential term accounts for the corresponding probability difference.
 
We have also identified the Galilean metric, introduced previously \cite{MKS}, in terms of a ``non-relativistic decomposition'' of the Minkowskian metric. It is interesting that the classical and quantum notions of non-relativistic spacetime seem to be different, {\it i.e.}, while a quantum particle can distinguish events differing by time coordinates of order $1/c^2$, a classical particle cannot.

\section*{Acknowledgments}

It is a pleasure to thank C. Chryssomalakos, E. Okon, D. Sudarsky and Y. Bonder for helpful discussions. This work has been partially supported by the Czech Ministry of Education, Youth and Sports within the project LC06002.

\end{document}